\documentclass[pra,showpacs,amsmath,amssymb,twocolumn]{revtex4}
\usepackage{bm,multirow}
\usepackage[usenames]{color}
\newcommand \beq{\begin{eqnarray}}
\newcommand \eeq{\end{eqnarray}}
\def\simge{\mathrel{%
       \rlap{\raise 0.511ex \hbox{$>$}}{\lower 0.511ex \hbox{$\sim$}}}}
\def\simle{\mathrel{
       \rlap{\raise 0.511ex \hbox{$<$}}{\lower 0.511ex \hbox{$\sim$}}}}
\usepackage{graphics}
\usepackage[dvips]{graphicx}
\usepackage{graphicx}
\newcommand{\qq}{\mathbf{q}}
\newcommand{\pp}{\mathbf{p}}
\newcommand{\kk}{\mathbf{k}}

\newcommand{\rr}{\mathbf{r}}

\newcommand{\be}{\begin{equation}}
\newcommand{\ee}{\end{equation}}
\newcommand{\bea}{\begin{eqnarray}}
\newcommand{\eea}{\end{eqnarray}}
\newcommand{\ba}{\begin{align}}
\newcommand{\ea}{\end{align}}

\begin{document}
\title{Short-range correlations in dilute atomic Fermi gases with spin-orbit coupling}
\author{Zhenhua Yu}
\affiliation{Institute for Advanced Study, Tsinghua University, Beijing 100084, China}

\date{\today}

\begin{abstract}
We study the short-range correlation strength of three dimensional spin half dilute atomic Fermi gases with spin-orbit coupling. The interatomic interaction is modeled by the contact pseudopotential. In the high temperature limit, we derive the expression for the second order virial expansion of the thermodynamic potential via the ladder diagrams. We further evaluate the second order virial expansion in the limit that the spin-orbit coupling constants are small, and find that the correlation strength between the fermions increases as the forth power of the spin-orbit coupling constants. At zero temperature, we consider the cases in which there are symmetric spin-orbit couplings in two or three directions. In such cases, there is always a two-body bound state of zero net momentum. In the limit that the average interparticle distance is much larger than the dimension of the two-body bound state, the system primarily consists of condensed bosonic molecules that fermions pair to form; we find that the correlation strength also becomes bigger compared to that in the absence of spin-orbit coupling. Our results indicate that generic spin-orbit coupling enhances the short-range correlations of the Fermi gases. Measurement of such enhancement by photoassociation experiment is also discussed.


\end{abstract}

\maketitle

\section{Introduction}

Recent experimental advances in generating synthetic gauge fields are motivated by simulating charged particles in solid state systems by neutral atoms \cite{dalibard,zhai}. In the presence of external magnetic fields, the degeneracy of a manifold of the atom's hyperfine spin states is lifted. The coupling between the manifold of the hyperfine spin states and external laser fields gives rise to dressed states. The adiabatic elimination of the high energy dressed states results in a low energy effective Hamiltonian in which synthetic gauge fields emerge. By such schemes, uniform vector potentials \cite{spielmanv}, synthetic magnetic \cite{spielmanm} and electric fields \cite{spielmane} are realized in condensates of $^{87}$Rb atoms. With the magnetic field's strength and the laser frequency fine tuned, a spin-orbit coupling bilinear in momentum and pseudo-spin operator components in one direction is engineered as well \cite{spielmanso}. The possibility of inducing spin-orbit couplings in two \cite{campbell} and three directions \cite{galitski} is further discussed theoretically. Similar attempts to synthesize gauge fields in atomic Fermi gases are under active experimental exploration \cite{zhang}.

It is an interesting question how the introduction of spin-orbit coupling would affect the correlations of dilute atomic gases.
In the BEC-BCS crossover problem, Tan noticed that the correlations in a homogeneous dilute two-component atomic Fermi gas have an asymptotic form \cite{tan}
\begin{align}
\langle\psi_\uparrow^\dagger(\rr)\psi_\downarrow^\dagger(0)\psi_\downarrow(0)\psi_\uparrow(\rr)\rangle=C\left(\frac1r-\frac1 {a_s}\right)^2,\label{asym}
\end{align}
in the regime where $r$ is much less than $d$ the mean distance between the particles  and bigger than $r_0$ the range of the interatomic interaction potential $U(r)$. Here $\psi_{\sigma}$ are the field operators for fermions and $a_s$ is the s-wave scattering length.
The correlation (or contact) strength at short distance $C$ has been shown to be linked with thermodynamic quantities through a series of remarkable relations \cite{tan, braaten, shizhong}, named as Tan's relations, one of which is
\begin{align}
C=-\frac m{4\pi}\frac{\partial f}{\partial a_s^{-1}},\label{dfda}
\end{align}
where $m$ is the mass of particles and the free energy density is $f=-T\log({\rm Tr} e^{-H/T})/V$ with $T$ the temperature and $V$ the volume of the system.

In the presence of spin-orbit coupling, Eqs.~(\ref{asym}) and (\ref{dfda}) should hold when $r_0$ is much smaller than the length scale corresponding to the spin-orbit coupling strength. For specific, let us consider the spin-orbit coupling of the form $h_{so}=\sum_{i=x,y,z}\kappa_i p_i\sigma_i$, where $\sigma_i$ are the Pauli matrices and $\kappa_i$ are the spin-orbit coupling constants. Originally, in the absence of $h_{so}$, Eq.~(\ref{asym}) can be derived from the observation \cite{shizhong} that given the scale separation $r_0\ll d$ in dilute Fermi gases, when one writes $\langle\psi_\uparrow^\dagger(\rr)\psi_\downarrow^\dagger(0)\psi_\downarrow(0)\psi_\uparrow(\rr)\rangle=C\chi^2(r)$, in the regime $r\lesssim r_0$, $\chi(r)$ satisfies the Schr\"odinger equation of the relative motion of two interacting fermions with zero energy
\begin{align}
\left[-\frac{\nabla^2}m+U(r)\right]\chi(r)=0.\label{chi}
\end{align}
The asymptotic form $\chi(r)\sim (1/r-1/a_s)$ for $r>r_0$ is required to connect with the behavior of $\chi(r)$ in the regime $r<r_0$ which is solely determined by $U(r)$; $a_s$ parameterizes the effects of $U(r)$ on $\chi(r)$. 
Note that since we assume that $U(r)$ is not fine tuned close to any resonance other than in the s-wave channel, the non s-wave parts of $\langle\psi_\uparrow^\dagger(\rr)\psi_\downarrow^\dagger(0)\psi_\downarrow(0)\psi_\uparrow(\rr)\rangle$ is neglected for $r\lesssim r_0$ due to the strong suppression by the centrifugal potential barrier.
The introduction of $h_{so}$ would modify Eq.~(\ref{chi}). Since experimental values of $\kappa_i$ are about $1/d$ \cite{spielmanso,zhang}, the inverse of the mean interparticle distance, we expect that the correction to $\chi$ due to $h_{so}$ is of order $\kappa_i r_0$, which is negligible in dilute gases. A two-body calculation using a square well model potential for equal spin-orbit couplings in three directions agrees with our expectation \cite{cui}. Given $\chi$ unchanged in the lowest order of $\kappa_i r_0$, in the same way as used in Refs.~\cite{shizhong, yue}, one can show that Eq.~(\ref{dfda}) stands. However, how spin-orbit coupling would affect the magnitude of the correlation strength at short distance is the problem that we are going to study below.

In this paper, we consider three dimensional spin half Fermi gases with spin-orbit coupling $h_{so}=\sum_{i=x,y,z}\kappa_i p_i\sigma_i$. The Hamiltonian of the system is
\begin{align}
H=&H_0+H_{int}+H_{so},\nonumber\\
H_0=&\int d^3\rr\nabla\Psi^\dagger(\rr)\nabla\Psi(\rr)/2m,\nonumber\\
H_{int}=& \bar g\int d^3\rr \psi^\dagger_\uparrow(\rr)\psi^\dagger_\downarrow(\rr)\psi_\downarrow(\rr)\psi_\uparrow(\rr),\nonumber\\
H_{so}=&\int d^3\rr\Psi^\dagger(\rr)h_{so}\Psi(\rr).
\end{align}
Here $\Psi=(\psi_\uparrow,\psi_\downarrow)^T$. The bare coupling constant $\bar g$ for the contact pseudopotential $U(\rr)=\bar g\delta(\rr)$ defined with the momentum cutoff $\Lambda$ is related to the s-wave scattering length $a_s$ via the renormalization
\begin{align}
\frac m{4\pi a_s}=\frac1 {\bar g}+\frac {m\Lambda}{2\pi^2}\label{renorm}.
\end{align}
We take $\hbar=1$ throughout. In the high temperature limit, we derive the second order virial expansion of the thermodynamic potential of the system by the ladder diagrams. We further evaluate the second order virial expansion perturbatively for small $\kappa_i$ and find that the correlation strength $C$ increases as the forth power of $\kappa_i$. At zero temperature, we consider two cases, equal spin-orbit couplings in two or three dimensions, in which there always exits a two-body bound state with zero net momentum no matter the value of $a_s$ \cite{shenoy}. In the limit that the mean interparticle distance is much larger than the size of the two-body bound state, by the fact that the leading contribution to the ground state energy comes from the binding energy of the bound state which fermions pair into, we show that $C$ becomes bigger compared to that in the absence of spin-orbit coupling. Our results indicate that nonzero spin-orbit coupling generically enhances the correlation strength of the Fermi gases. Such enhancement can be detected in photoassociation experiment.

\section{Virial expansion and ladder diagrams}
When the temperature $T$ is high, the grand canonical partition function, $\mathcal Z={\rm Tr}e^{-(H-\mu N)/T}$ with $\mu$ the chemical potential, can be approximated by a virial expansion in terms of the fugacity $\eta=e^{\mu/T}$ which is a small number. The effects of pairwise interactions first appear in the second order virial expansion.
For two component Fermi gases interacting through a short range central potential without spin-orbital coupling, the second virial coefficient due to interactions has been derived \cite{beth}
\begin{align}
b_2=\sum_n e^{|E_n|/T}+\sum_\ell \int_0^\infty \frac{dk}\pi (2\ell+1) \frac{d\delta_\ell(k)}{dk} e^{-k^2/mT}\label{b2},
\end{align}
where $E_n$ are the binding energies of two-body bound states and $\delta_{\ell}$ are the phase shifts in the $\ell$th partial waves.
In the presence of spin-orbital coupling, $h_{so}$ couples scatterings in different partial waves to each other; this coupling renders classification in terms of angular momentum as in Eq.~(\ref{b2}) impossible. However, since the ladder diagrams exhaust the two-body scattering processes, we use them to calculate the second order virial expansion \cite{randeria,pethick}.

It is instructive to demonstrate how the ladder diagrams reproduce a result agreeable with Eq.~(\ref{b2}) for a Fermi gas whose Hamiltonian is $H_0+H_{int}$. The variation of the thermodynamic potential $\Omega=-T\log\mathcal Z$ due to the ladder diagrams is \cite{pethick}
\begin{align}
\delta\Omega=\sum_{\mathbf P}\int_{\mathcal C}\frac{d\zeta}{2\pi i}f_B(\zeta)\log\left(\frac m{4\pi a_s}-\Pi(P,\zeta)\right),\label{do}
\end{align}
with
\begin{align}
\Pi(P,\zeta)=\int\frac{d^3\mathbf q}{(2\pi)^3}
\left\{\frac{1-f(\xi_{\mathbf P/2+\mathbf q})-f(\xi_{\mathbf P/2-\mathbf q})}{\zeta-\xi_{\mathbf P/2+\mathbf q}-\xi_{\mathbf P/2-\mathbf q}}+\frac1{\epsilon_{\mathbf q}}\right\}.
\end{align}
Here $\xi_{\mathbf q}=\epsilon_{\mathbf q}-\mu$, $\epsilon_{\mathbf q}=q^2/2m$, $f_B$ and $f$ are the Bose and Fermi distribution functions respectively. The renormalization Eq.~(\ref{renorm}) has been used to obtain Eq.~(\ref{do}). The branch cut of the logarithmic function lies on the positive real axis of its argument. The contour $\mathcal C$ wraps the real axis of the integral variable $\zeta$.

To order of $\eta^2$, we neglect the Fermi distribution functions in $\Pi$, since they contribute at least an extra factor of $\eta$; the argument of the logarithmic function in Eq.~(\ref{do}) becomes the inverse of the T-matrix in vacuum $t^{-1}(\zeta-P^2/4m+2\mu)$, where
\begin{align}
t(\zeta)=\left[\frac m{4\pi a_s}+\frac{i m^{3/2}}{4\pi}\zeta^{1/2}\right]^{-1}.
\end{align}
After changing the variable $\zeta'=\zeta-P^2/4m+2\mu$, we have
\begin{align}
\delta\Omega=\sum_{\mathbf P}\int_{\mathcal C}\frac{d\zeta'}{2\pi i}f_B(\zeta'+P^2/4m+2\mu)\log\left(t^{-1}(\zeta')\right).
\end{align}
It is generically true that the singularities of $\log(t^{-1}(P,\zeta'))$ on the real axis of $\zeta'$ are left bounded.
We deform the contour $\mathcal C$ to wrap the part of the real axis of $\zeta'$ right to the most left singularity. On this contour $\mathcal C$,
we expand $f_B(\zeta'+P^2/4m+2\mu)$ in the integrand to the lowest order of $\eta$, which is of order $\eta^2$, as
\begin{align}
\delta\Omega\approx \eta^2\sum_{\mathbf P}e^{-P^2/4mT}\int_{\mathcal C}\frac{d\zeta'}{2\pi i}e^{-\zeta'/T}\log\left(t^{-1}(\zeta')\right).
\end{align}
Direct evaluation of the above equation gives
\begin{align}
\delta\Omega^{(0)}=-\frac{2^{3/2}\eta^2 T V}{\lambda^3}\tilde b_2(\lambda/a_s),\label{do0}
\end{align}
with
\begin{align}
&\tilde b_2(\lambda/a_s)\nonumber\\
&=\theta(a_s)e^{\lambda^2/2\pi a_s^2}+\int_0^\infty \frac{dk}{\pi}\frac{(-a_s)}{1+(ka_s)^2}e^{- k^2\lambda^2/2\pi}\nonumber\\
&=\frac12[1+{\rm Erf}(\lambda/\sqrt{2\pi}a_s)]e^{\lambda^2/2\pi a_s^2}.\label{tb2}
\end{align}
Here the thermal wavelength is $\lambda\equiv\sqrt{2\pi/mT}$. Equation (\ref{tb2}) has been obtained by calculating the partition function from the two-body eigenenergies in Refs.~\cite{ho,yue}. Given that the contact pseudopotential $\bar g\delta(\rr)$ scatters only the s-wave, and $\cot\delta_s(k)=-1/ka_s$, and $-1/ma_s^2$ is the binding energy for the only bound state when $a_s>0$, Eq.~(\ref{tb2}) agrees with Eq.~(\ref{b2}).
For fixed density $n=N/V$ with $N$ the total number of fermions, Eq.~(\ref{do0}) is
\begin{align}
\delta\Omega^{(0)}=-2^{-1/2}n^2 T V\lambda^3\tilde b_2(\lambda/a_s)
\end{align}
since $n=2\eta/\lambda^3$ to order of $\eta$,

In the presence of $H_{so}$, $\delta\Omega$ retains the form of Eq.~(\ref{do}) with $\Pi$ replaced by 
\begin{align}
&\Pi_{so}(\mathbf P, \zeta)\nonumber\\
&=\frac1{4}\int\frac{d^3\mathbf q}{(2\pi)^3}\left(
\sum_{\alpha,\alpha'=\pm1}\frac{1-\alpha\alpha'\phi_{\mathbf P,\mathbf q}}{\zeta-E^{\alpha}_{\mathbf P/2+\mathbf q}-E^{\alpha'}_{\mathbf P/2-\mathbf q}}
+\frac4{\epsilon_{\mathbf q}}\right),\label{pisoc}
\end{align}
with $E^{\alpha}_{\mathbf q}=\xi_{\mathbf q}+\alpha\Delta_{\mathbf q}$, $\Delta_{\mathbf q}=\sqrt{\sum_{i}\kappa_i^2 q_i^2}$, and $\phi_{\mathbf P,\mathbf q}=[\sum_{i}\kappa_i^2(P_i^2/4-q_i^2)]/\Delta_{\mathbf P/2+\qq}\Delta_{\mathbf P/2-\qq}$. The index $\alpha=\pm1$ picks up different helicity branches of noninteracting particles \cite{zhai}. Note that Fermi distribution functions have been neglected in $\Pi_{so}$ for the same reason as stated before.

\section{Perturbation in high temperature limit}
While to obtain the behavior of the second order virial expansion $\delta\Omega$ as a function of arbitrary values of $\kappa_i$ requires a full evaluation of the multi-dimensional integral (cf.~Eqs.~(\ref{do}) and (\ref{pisoc})), in the following, we calculate $\delta\Omega$ perturbatively in terms of $m\lambda\kappa_i$ in the high $T$ limit in which $\lambda\to0$. Since the integrand of $\Pi_{so}$ is invariant under the transformation $\kappa_i\to-\kappa_i$, the perturbation series of $\delta\Omega$ consists of only even orders of $m\kappa_i\lambda$. To the forth order of $\kappa_i$
\begin{align}
&\frac m{4\pi a_s}-\Pi_{so}(\mathbf P, \zeta'+P^2/4m-2\mu)=t^{-1}(\zeta')-4\int\frac{d^3\mathbf q}{(2\pi)^3}\nonumber\\
&\times\left[\frac{\sum_i\kappa_i^2q_i^2}{(\zeta'-q^2/m)^3}+\frac{\sum_{i,j}\kappa_i^2\kappa_j^2(4q_i^2q_j^2+p_i^2q_j^2-p_ip_jq_iq_j)}{(\zeta'-q^2/m)^5}\right].\label{dosoc}
\end{align}

For the second order virial expansion $\delta\Omega$ to second order of $\kappa_i$,
\begin{align}
&\delta\Omega^{(2)}\nonumber\\
&=\eta^2\left(\sum_i\kappa_i^2\right)\sum_{\mathbf P}\int_{\mathcal C}\frac{d\zeta'}{2\pi i}e^{-(\zeta'+P^2/4m)/T}\frac{im^3}{8\pi\sqrt{m\zeta'}}t(\zeta')\nonumber\\
&=-m\left(\sum_{i}\kappa_i^2\right)\frac {2^{3/2}\eta^2 V}{\lambda^3}\tilde b_2(\lambda/a_s)\nonumber\\
&=\left[\sum_i(m\lambda\kappa_i)^2/2\pi\right]\delta\Omega^{(0)}.\label{do2}
\end{align}

\begin{figure}
\includegraphics[width=3.3in]{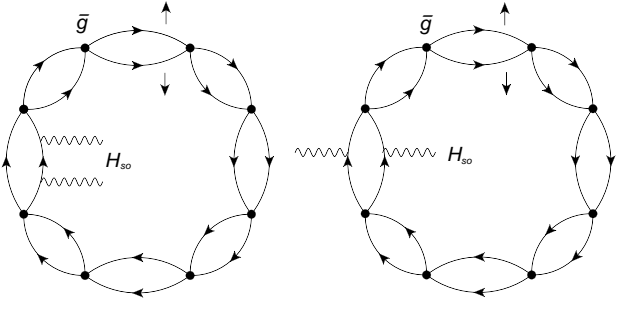}\caption{The Ladder diagrams contributing to the variation of the thermodynamical potential of the oder of the square of the spin-orbit coupling. The point vertices are $\bar g$, which are connected by a pair of free particle propagators of $H_0$. The wiggling vertices correspond to $H_{so}$. }\label{fey}
\end{figure}

Equation~(\ref{do2}) can be reproduced from the diagrams shown in Fig.~(\ref{fey}).
The perturbation $\delta\Omega$ by $H_{so}$ comes from the ladder diagrams \cite{nsr} with the vertex $H_{so}$ attached.
Since the system in the absence of $H_{so}$ are invariant under the reflection of momentum $\pp\to-\pp$, the diagrams with a single vertex $\sum_i\kappa_ip_i\sigma_i$ attached must be identically zero. Of second order of $\kappa_i$, similarly, the diagrams proportional to $\kappa_i\kappa_j$ vanish if $i\neq j$; for $i=j$,  two $\kappa_i p_i \sigma_i$ must attach to the same pair of free particle propagators as shown in Fig.~(\ref{fey}). To see this point, given that the three directions are equivalent, let us consider attaching $\kappa_x p_x \sigma_x$ to the ladder diagrams. We choose the free particle propagators diagonalizing the $z$ component of the spin operator $\sigma_z$. Since the interactions have $SU(2)$ symmetry and $\kappa_x p_x\sigma_x$ flips the spin by unity, diagrams in Fig.~(\ref{fey}) are the only nonzero ones contributing to $\Delta\Omega^{(2)}$ within the ladder diagrams.
The class of the diagrams represented by the left one in Fig.~(\ref{fey}) gives
\begin{align}
\delta\Omega^{(2)}_l=&\frac{T^2}{V}\sum_{\mathbf P,\mathbf q,\zeta,z,i}\left[\frac{\kappa_i^2(P_i/2+q_i)^2}{(\zeta-z-\xi_{\mathbf P-\qq})(z-\xi_{\mathbf P+\qq})^3}\right.\nonumber\\
&\left.+\frac{\kappa_i^2(P_i/2-q_i)^2}{(\zeta-z-\xi_{\mathbf P+\qq})(z-\xi_{\mathbf P-\qq})^3}\right]t(\zeta-P^2/4m+2\mu),\label{dop}
\end{align}
and the class by the right gives
\begin{align}
\delta\Omega^{(2)}_r=&-\frac{T^2}{ V}\sum_{\mathbf P,\mathbf q,\zeta,z,i}\frac{\kappa_i^2(P^2_i/4-q^2_i)}{(\zeta-z-\xi_{\mathbf P-\qq})^2(z-\xi_{\mathbf P+\qq})^2}\nonumber\\
&\times t(\zeta-P^2/4m+2\mu).\label{dopp}
\end{align}
Here $\zeta$ and $z$ are the bosonic and fermionic Matsubara frequencies respectively. Of order $\eta^2$, the sum of Eqs.~(\ref{dop}) and (\ref{dopp}) equals Eq.~(\ref{do2}).

For fixed density $n$, since $n=2\eta[1+\sum_i(m\lambda\kappa_i)^2/4\pi]/\lambda^3$ to first order of $\eta$ and second order of $\kappa_i$, $\delta\Omega^{(0)}+\delta\Omega^{(2)}=-2^{-1/2}TV\lambda^3 n^2 \tilde b_2(\lambda/a_s)$; the second virial coefficient is still $\tilde b_2(\lambda/a_s)$, not changed to the order of $\kappa_i^2$. Consequently, the correlation strength $C$ does not change as well. This absence of modification of $C$ can be understood in the following way. The contributions to $\delta\Omega$ proportional to $\kappa_i^2$ from the three directions are additive and of the same form. We can reproduce the coefficient of $\kappa_z^2$ in Eq.~(\ref{do2}) by introducing the perturbation Hamiltonian $H_p=\int d^3\rr \Psi(\rr)\kappa_z p_z \sigma_z \Psi(\rr)$ to $H_0+H_{int}$. Note that the unitary tranformation $U=\exp[im\kappa_z\int d^3\rr z (\psi^\dagger_\uparrow(\rr)\psi_\uparrow(\rr)-\psi^\dagger_\downarrow(\rr)\psi_\downarrow(\rr))]$ transforms $K=H_0+H_{int}+H_p-\mu N$ as $UKU^\dagger=K-(m\kappa_z^2/2)N$. We have
\begin{align}
\Omega(T,\mu,\kappa_z)=-T\log({\rm Tr}e^{-K/T})=\Omega(T,\mu+m\kappa_z^2/2,0).\label{shift}
\end{align}
For the second order virial expansion $\delta\Omega$ of order $\kappa_z^2$,
\begin{align}
\delta\Omega(T,\mu,\kappa_z)&=\delta\Omega(T,\mu,0)+\frac{\partial \delta\Omega(T,\mu,0)}{\partial\mu}(m\kappa_z^2/2)\nonumber\\
&=\delta\Omega^{(0)}[1+(m\lambda\kappa_z)^2/2\pi],
\end{align}
which agrees with Eq.~(\ref{do2}).  
For the correlation strength $C$, let us write its correction of second order $\kappa_i$ as $A\sum_i(m\lambda\kappa_i)^2$. In the case of with spin-orbit coupling only in $z$ direction, $\delta C=A(m\lambda\kappa_z)^2$. However, according to Eq.~(\ref{shift}), the effect of the spin-orbit coupling only in one direction is equivalent to shifting $\mu$.
Physically, for fixed density, a chemical potential shift should not affect $C$ at all; one concludes $A=0$.

The effects of the spin-orbit coupling on $C$ can be revealed by calculating $\delta\Omega$ to the forth power of $\kappa_i$. From Eq.~(\ref{dosoc}), the non-cross terms $\propto \kappa_i^4$ give
\begin{align}
\delta\Omega^{(4)}_{nc}
=&-{\eta}^2\left(\sum_i\kappa_i^4\right)\frac{2^{3/2}V}{\lambda^3}\int_{\mathcal C}\frac{d\zeta'}{2\pi i}e^{-\zeta'/T}\nonumber\\
&\times\left[\frac{im^{7/2}}{32\pi \zeta'^{3/2}}t(\zeta')-\frac{m^5}{128\pi^2\zeta'}t^2(\zeta')\right]\nonumber\\
=&\sum_i(m\lambda\kappa_i^2)^4\delta\Omega^{(0)}/8\pi^2,
\end{align}
a result expected from expanding Eq.~(\ref{shift}) in terms of $\kappa_z^2$.
The cross terms $\propto\kappa_i^2\kappa_j^2$ for $i\neq j$ are
\begin{align}
&\delta\Omega^{(4)}_c
={\eta}^2\left(\sum_{i<j}\kappa_i^2\kappa_j^2\right)\frac{2^{3/2}V}{\lambda^3}\frac{m^5}{16\pi}\int_{\mathcal C}\frac{d\zeta'}{2\pi i}e^{-\zeta'/T} \nonumber\\
&\times\left\{it(\zeta')\left[\frac{mT}2\frac1{(m\zeta')^{5/2}}
-\frac13\frac1{(m\zeta')^{3/2}}\right]+\frac1{4\pi\zeta'}t^2(\zeta')\right\}.
\end{align}
After changing the integral variable $z=2\pi\zeta'/T$ and integrating by part, we have
\begin{align}
\delta\Omega^{(4)}_c=&{\eta}^2\left(\sum_{i<j}\kappa_i^2\kappa_j^2\right)\frac{2^{3/2}V}{\lambda^3}\frac{m^3\lambda^2}{4}\int_{\mathcal C}\frac{dz}{2\pi i} e^{-z/2\pi}\nonumber\\
&\times\left[
-\frac\pi{z}\frac1{(\alpha+i\sqrt{z})^4}+\frac i 3\frac1{\sqrt{z}}\frac1{(\alpha+i\sqrt{z})^3}\right.\nonumber\\
&\left.+\frac1{6z}\frac1{(\alpha+i\sqrt{z})^2}
+\frac{2i}{3\pi}\frac1{\sqrt{z}}\frac1{(\alpha+i\sqrt{z})}\right],
\end{align}with $\alpha=\lambda/a_s$.

\begin{figure}
\includegraphics[width=3in]{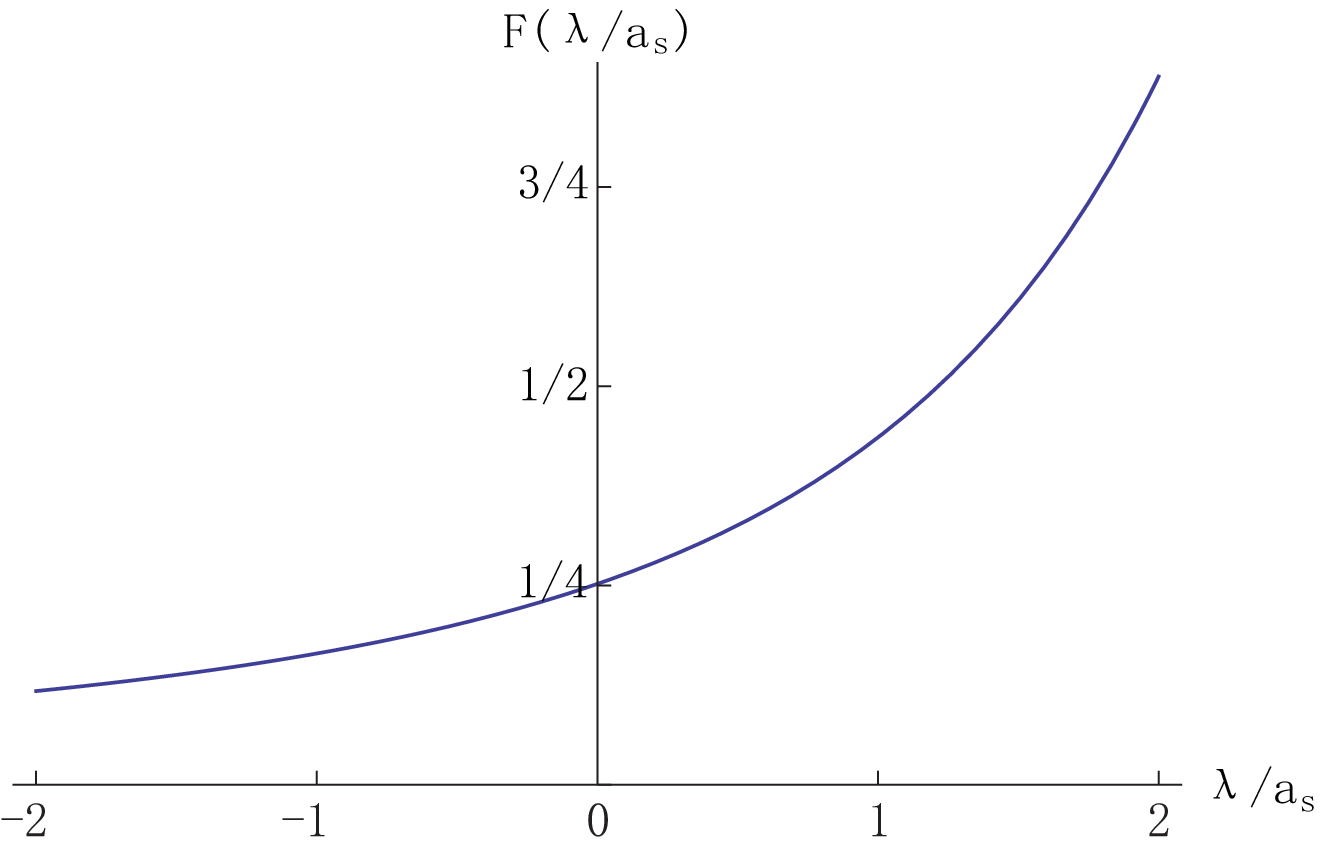}\caption{(Color online) $F(\lambda/a_s)$ vs $\lambda/a_s$ calculted from Eq.~(\ref{falpha}).}\label{fx}
\end{figure}

Direct evaluation of the contour integral yields
\begin{align}
\delta\Omega^{(4)}_c=&-\frac{2^{3/2}\eta^2TV}{\lambda^3}\frac{F(\lambda/a_s)}{8\pi}\left(\sum_{i<j}m^4\lambda^4\kappa_i^2\kappa_j^2\right)
\end{align}
where
\begin{align}
F(\alpha)=&e^{\alpha^2/2\pi}\left(\frac\pi{\alpha^4}-\frac2{3\alpha^2}+\frac1\pi\right)\left[1+{\rm Erf}(\alpha/\sqrt{2\pi})\right]\nonumber\\
&
-\frac\pi{\alpha^4}+\frac1{6\alpha^2}-\sqrt2\left(\frac1{\alpha^3}-\frac1{3\pi\alpha}\right).\label{falpha}
\end{align}
Figure (\ref{fx}) shows $F(\lambda/a_s)$ as a function of $\lambda/a_s$, positive and well-behaved everywhere.

\begin{figure}
\includegraphics[width=3.2in]{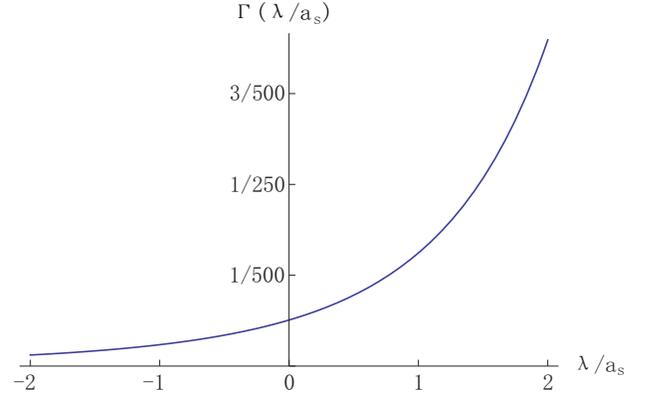}\caption{(Color online) $\Gamma(\lambda/a_s)$ vs $\lambda/a_s$ calculated from Eq.~(\ref{gammaalpha}).}\label{gammax}
\end{figure}

To the same order of $\kappa_i$
\begin{align}
n=&\frac{2\eta}{\lambda^3}\left\{1+\frac{m^2\lambda^2}{4\pi}\left(\sum_i\kappa_i^2\right)
+\frac{m^4\lambda^4}{16\pi^2}\left[\frac12\sum_i\kappa_i^4\right.\right.\nonumber\\
&\left.\left.+\frac13\left(\sum_{i<j}\kappa_i^2\kappa_j^2\right)\right]\right\},
\end{align}
the thermodynamical potential variation in terms of $n$ is
\begin{align}
\delta\Omega=&-2^{-1/2}TV\lambda^3n^2\left[\tilde b_2(\lambda/a_s)+\left(\frac{F(\lambda/a_s)}{8\pi}\right.\right.\nonumber\\
&\left.\left.-\frac{\tilde b_2(\lambda/a_s)}{6\pi^2}\right)\left(\sum_{i<j}m^4\lambda^4\kappa_i^2\kappa_j^2\right)\right].
\end{align}
By Eq.~(\ref{dfda}) the variation of $C$ is
\begin{align}
\delta C=2^{-3/2}\lambda^2 n^2 \left(\sum_{i<j}m^4\lambda^4\kappa_i^2\kappa_j^2\right)
\Gamma(\lambda/a_s)
\end{align}
with
\begin{align}
\Gamma(\lambda/a_s)=\frac{\partial}{\partial (\lambda/a_s)}\left(\frac{F(\lambda/a_s)}{8\pi}-\frac{\tilde b_2(\lambda/a_s)}{6\pi^2}\right).\label{gammaalpha}
\end{align}
Figure~(\ref{gammax}) shows that $\Gamma$ is always positive; the correlation strength increases as the forth power of $\kappa_i$ in the limit $m\lambda\kappa_i\to0$. Note that since $\delta C\propto\sum_{i<j}\kappa_i^2\kappa_j^2$, the correlation strength changes only if spin-orbit coupling constants are nonzero at least in two directions.

\section{Zero temperature}
At zero temperature, we consider that there is a symmetric spin-orbit coupling in either two or three directions, i.e.~, $\kappa_x=\kappa_y=\kappa$ and $\kappa_z=0$, or $\kappa_x=\kappa_y=\kappa_z=\kappa$. It has been shown that for the two cases there is always a two-body bound state with zero center of mass momentum for all $a_s$ \cite{shenoy}. The eigenenergy of this bound state $\epsilon$ is given by
\begin{align}
\frac m{4\pi a_s}-\Pi_{so}(0,\epsilon)=0,\label{eb}
\end{align}
where in the expression of $\Pi_{so}$ the chemical potential is set zero. In the limit that the mean distance between particles $d\sim n^{-1/3}$ is much larger than the size of the bound state $\ell_b$, the system can be considered as primarily consisting of bosonic molecules which two fermions pair into. These molecules condense into the zero momentum state. Thus we expand in terms of the small number $\ell_b n^{1/3}$ and have the leading contribution to the energy of the system coming from the binding energy of the molecules as $E=N \epsilon/2$.

For $\kappa_x=\kappa_y=\kappa$ and $\kappa_z=0$, Eq.~(\ref{eb}) reduces to \cite{shenoy}
\begin{align}
\frac1{m\kappa a_s}=&\sqrt{1+\frac{E_b}{m\kappa^2}}-\log\left(1+\sqrt{1+\frac{E_b}{m\kappa^2}}\right)\nonumber\\
&-\frac12\log\left(\frac{m\kappa^2}{E_b}\right).\label{eb2}
\end{align}
The binding energy $E_b$ is defined with respect to the scattering threshold energy $E_{th}=-m\kappa^2$ as $E_b\equiv |\epsilon-E_{th}|$. The wavefunction of the bound state is \cite{shenoy}
\begin{align}
&\psi_b(\rr)\nonumber\\
&=\frac1{\sqrt{\mathcal N}}\left(\psi_s(\rr)|\uparrow\downarrow-\downarrow\uparrow\rangle+\psi_a(\rr)|\uparrow\uparrow\rangle+\psi^*_a(\rr)|\downarrow\downarrow\rangle\right),
\end{align}
where $\psi_s(\rr)=-\sum_{\kk,\alpha}\cos(\kk\cdot\rr)/[2(\epsilon_{\kk}+\alpha\Delta_\kk)-\epsilon]$, $\psi_a(\rr)=i\sum_{\kk,\alpha}\alpha e^{-i\phi_\kk}\sin(\kk\cdot\rr)/[2(\epsilon_{\kk}+\alpha\Delta_\kk)-\epsilon]$ with $\phi_\kk$ the azimuthal angle of $\kk$, and the normalization factor $\mathcal N=\sum_{\kk,\alpha}2V/[2(\epsilon_{\kk}+\alpha\Delta_\kk)-\epsilon]^2$.

It is instructive to check that $C$ extracted from the correlation at short distance through Eq.~(\ref{asym}) satisfies Eq.~(\ref{dfda}). By the wavefunction $\psi_b$,
\begin{align}
C&=\lim_{r\to0}r^2\langle\psi^\dagger_\uparrow(\rr)\psi^\dagger_\downarrow(0)\psi_\downarrow(0)\psi_\uparrow(\rr)\rangle\nonumber\\
&=(N/V) \lim_{r\to0}r^2 \mathcal N^{-1} |\psi_s(\rr)|^2\nonumber\\
&=(4\pi^2)^{-1}\mathcal N^{-1} m^2 VN.\label{2dc}
\end{align}
On the other hand, from Eq.~(\ref{eb})
\begin{align}
-\frac m{4\pi}\frac{\partial (E/V)}{\partial (1/a_s)}&=\frac{m^2 N}{(4\pi)^2 V}\left[\frac1{2V}\sum_{\kk,\alpha}\frac1{[\epsilon-2(\epsilon_{\kk}+\alpha\Delta_\kk)]^2}\right]^{-1},
\end{align}
which matches Eq.~(\ref{2dc}).

In the case of $\kappa_x=\kappa_y=\kappa_z=\kappa$, from Eq.~(\ref{eb}) the binding energy $E_b$ is given by \cite{shenoy}
\begin{align}
mE_b=\frac14\left(\frac1{a_s}+\sqrt{\frac1{a_s^2}+4m^2\kappa^2}\right)^2;\label{eb3}
\end{align}
the bound state wavefunction is
\begin{align}
\psi_b(\rr)
=&\frac1{\sqrt{\mathcal N}}\left[\frac{e^{-\sqrt{m E_b}r}}r\left(\frac\kappa{\sqrt{E_b/m}}\sin(m\kappa r)+\cos(m\kappa r)\right)\right.\nonumber\\
&\times|\uparrow\downarrow-\downarrow\uparrow\rangle
+i\left((\sqrt{m E_b}+1/r)\sin(m\kappa r)\right.\nonumber\\
&\left.\left.-m\kappa\cos(m\kappa r)\right)\frac{e^{-\sqrt{m E_b}r}}{\sqrt{m E_b}r}|\uparrow\downarrow+\downarrow\uparrow\rangle_{\hat\rr}\right],
\end{align}
where the subscript means that $\hat\rr$ is the spin quantization axis. The normalization factor is $\mathcal N = 2\pi[m^2\kappa^2/(mE_b)^{3/2}+1/(mE_b)^{1/2}]$. It is also straightforward to show that $C$ extracted from the correlation function maintains Eq.~(\ref{dfda}).

\begin{figure}
\includegraphics[width=3.2in]{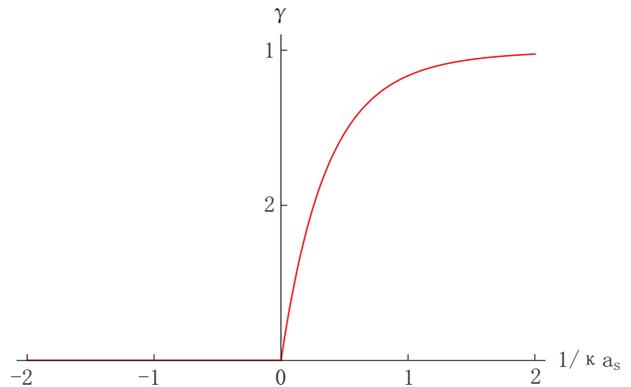}\caption{(Color online) The ratio $\gamma$ vs $1/\kappa a_s$.}\label{ga}
\end{figure}

In the absence of spin-orbit couplings, $C=n/4\pi a_s$ in the BEC limit $n^{1/3}a_s\to 0^+$. The correlation strength decreases as $a_s$ increases and acquires a universal value $C\approx 2.7\times (3\pi^2 n)^{4/3}/36\pi^4$ at unitarity $1/a_s=0$. In the BCS limit $n^{1/3}a_s\to 0^-$, $C=a_s^2n^2/4$. For the two cases considered above, spin-orbit coupling ensures the existence of the two-body bound state of zero net momentum. In the limit $\ell_b n^{1/3}\to0$, $C$ is contributed primarily from the bound state that the fermions pair into, and thus is proportional to the density $n$. In the limit $\kappa a_s\to 0^+$, for symmetric 2D couplings, $E_b=1/ma_s^2+m\kappa^2-m^3\kappa^4a_s^2/3$ and $C=(n/4\pi a_s)(1+m^4\kappa^4 a_s^4/6)$; for symmetric 3D couplings, $E_b=1/ma_s^2+2m\kappa^2-m^3 \kappa^4 a_s^2$ and $C=(n/4\pi a_s)(1+m^4\kappa^4 a_s^4/2)$. The variation of $C$ proportional to $\kappa^4$ and ratio between the coefficients of the $\kappa^4$ terms is expected from perturbative calculations as done before at high temperatures. Generally, combining Eqs.~(\ref{dfda}), (\ref{eb2}) and (\ref{eb3}), we have
\begin{align}
&\lim_{n\to0}\frac{ C(\kappa_x=\kappa_y=\kappa_z=0)}{C(\kappa_x=\kappa_y=\kappa_z=\kappa)}\nonumber\\
&=\begin{cases} \frac4{2+\sqrt{1+4m^2\kappa^2a_s^2}+1/\sqrt{1+4m^2\kappa^2a_s^2}}<1, & \text{for $a_s>0$;}\\
0, & \text{for $a_s<0$.}\end{cases}
\end{align}
Figure (\ref{ga}) plots $\gamma=\lim_{n\to0}C(\kappa_x=\kappa_y=\kappa_z=0)/C(\kappa_x=\kappa_y=\kappa,\kappa_z=0)$ versus $1/\kappa a_s$. Thus $C$ increases in the presence of symmetric spin-orbit coupling in two or three directions.

\section{Discussion}
Our results at high temperature and at zero temperature indicate that generic spin-orbit couplings enhances the correlation strength $C$. This enhancement is the net result of the two-fold effects brought about by $h_{so}$: the change of the density of states of noninteracting particles $\rho(\omega)$ and the mixing of scattering in different channels. The former effect is clearly reflected in determining the zero net momentum two-body bound state with symmetric spin-orbit coupling in two or three directions.
For the two cases, Eq.~(\ref{eb}) involves only the scattering between two particles of the same helicity and can be written as
\begin{align}
\frac m{4\pi a_s}-\int_0^{\infty}d\omega\left(\frac{\rho(\omega)}{\epsilon+m\kappa^2-2\omega}+\frac m{2\pi^2}\sqrt{\frac m{2\omega}}\right)=0.
\end{align}
The change of $\rho(\omega)$, especially in the limit $\omega\to0$ from $\sim\sqrt\omega$ to $\sim{\rm const.}$ and $\sim 1/\sqrt\omega$ respectively, gives rise to the two-body bound state no matter the value of $a_s$. The existence of this bound state guarantees the increase of $C$ in the low density limit at zero temperature. In the situation where the scattering between two particles with nonzero net momentum shall be taken into account, scatterings in intra- and inter-helicity channels are coupled together. This mixture can be seen by expressing $H_{int}$ in terms of the fermion operators diagonalizing $H_0+H_{so}$ \cite{baym}. Terms in $\Pi_{so}$ which determines the second order virial expansion correspond to different channels. Recently Ref.~\cite{bcs} employ the BCS mean field theory to study the ground state of attractive spin half fermions with spin-orbit coupling, and find that the BCS pairing gap $\Delta$ increases with spin-orbit coupling constants. This finding agrees with ours since within the BCS theory, $C=-(m/4\pi)[\partial (E/V)/\partial a_s^{-1}]\sim \Delta^2$.

The enhancement of $C$ in Fermi gases with spin-orbit coupling can be measured by photoassociation experiment. Previous experiment \cite{hulet}, across the Feshbach resonance between the lowest two hyperfine states of $^6$Li at magnetic field $834$G, associates a pair of Fermi atoms in the closed channel of the Feshbach resonance into a molecular state. The resultant molecules lose from the trap confining the Fermi gas. The loss rate of atoms $R$ has been shown to be proportional to $C$ \cite{shizhong, yue, castin}. In the presence of spin-orbit coupling, the relation between the loss rate $R$ and the correlation strength $C$ should remain intact since the photoassociation only involves physical processes at a distance $\sim r_0$ much shorter than the scale introduced by the spin-orbit coupling constants. The change of $C$ shall be clearly reflected in the atom loss rate.

\section*{Acknowledgement}
We thank Shizhong Zhang for bringing Ref.~\cite{randeria} to our attention. We are grateful to Ran Qi and Zeng-Qiang Yu for extensive and instructive discussions. We achnowledge Tin-Lun Ho, Hui Zhai and Xiaoling Cui for critical reading of the manuscript. This work is supported in part by Tsinghua University Initiative Scientific
Research Program, and NSFC under Grant No.~11104157.

\end{document}